\documentclass{eas}
\usepackage{graphicx}
\begin{document}

%%-----------------------------
%%      the top matter
%%-----------------------------
\title{The Vertical Structure of the Halo Rotation} 
\author{T.D. Kinman}\address{NOAO, P.O. Box 26732, Tucson, Arizona 85726-6732, USA \email{kinman@noao.edu}}
\secondaddress{
NOAO is operated by the Association of Universities for research in
Astronomy, Inc., under contract with the National Science Foundation}
\author{C. Cacciari}\address{INAF, Osservatorio Astronomico, Bologna, Italy}
\author{A. Bragaglia}\sameaddress{3}
\author{A. Buzzoni}\address{INAF, Centro Galileo Galilei, La Palma, Spain}
\author{A. Spagna}\address{INAF, Osservatorio Astronomico, Torino, Italy}
\begin{abstract}
New GSC-II proper motions of RR Lyrae and Blue Horizontal Branch (BHB) stars 
near the North Galactic Pole are used to show that the Galactic Halo 5 kpc above
the Plane has a significantly retrograde galactic rotation.
\end{abstract}
\maketitle
%%-----------------------------
%%      your text
%%-----------------------------
\section{Introduction}
Recent determinations of the mean rotation ($<$V$>$) of the field component of
the Galactic halo are 
summarized in Table 1 for stars within 2 kpc of the Sun and in Table 2 for
stars at distances (Z) more than 4 kpc above the Plane. 
Metal-poor subdwarfs are mostly discovered by their high proper motions 
and so a substantial correction for kinematic bias is required if they are to
be used as tracers. No such kinematic bias correction is needed for RR Lyrae 
stars or 
halo K giants (discovered from objective prism surveys) but (as also for the 
subdwarfs) a correction is needed for a thick disk component. The first four
estimates in Table 1 use somewhat different approaches to this disk correction 
and all give values of $<$V$>$ for the halo in the solar neighbourhood that are 
close to that of the local circular velocity of
$-$220 km/s (Kerr \& Lynden-Bell \cite{K86}); so there is some 
consensus that the local halo does not rotate. 

The situation is different for estimates of 
$<$V$>$ for halo stars out of the Plane (Table 2). The only estimate for
an {\it in situ} out-of-plane sample is that of Majewski (\cite{M92}) and 
Majewski \etal~(\cite{M96}) who found $<$V$>$ to be 55 km/s retrograde. 
The estimates of Carney (\cite{C99}) and Chiba \& Beers (\cite{C00}) come
from stars whose calculated orbits take them more than 4 kpc from the plane;
they find no rotation or a slightly prograde rotation for these stars. The
present work attempts to resolve this discrepancy; it grew from earlier studies
of halo stars in the North Galactic Cap (Kinman \etal~\cite{K96}) which
confirmed the streaming motion (in the W vector) that Majewski found for his 
subdwarf sample in SA 57.

The determination of the rotation vector V for stars in the North Galactic Cap 
depends critically on selecting appropriate tracer halo stars and having 
accurate distances and proper motions. 
The recent availability of
absolute proper motions (Spagna \etal, \cite{S96}) based on the GSC-II  
catalogue (Lasker \etal, \cite{L95}; McLean \etal, \cite{M00}) 
 affords a new opportunity to evaluate the rotation 
of the halo outside the Plane. This paper gives preliminary results from a 
limited sample.

\begin{table*}
{{\bf Table~1.}The Halo rotation from nearby stars (Z $\leq$ 2 kpc) } 
\begin{flushleft}
\begin{tabular}{lccc} 
\noalign{\smallskip}
\hline 
Halo tracer  & No. of stars& Mean Rotation & Reference  \\
(Stellar type)& in sample   &  $<$V$>$ (km/s)       &            \\
\noalign{\smallskip}
\hline 
\noalign{\smallskip}
RR Lyrae          & 84 & $-$219$\pm$10  & Martin \& Morrison(\cite{M98})    \\
RR Lyrae          &162 & $-$210$\pm$12  & Layden \etal~(\cite{L96})        \\
RR Lyrae & 124 & $-$217$\pm$21 & Chiba \& Yoshi (\cite{C98}) \\
\& K giant$^{a}$ & &           &                             \\
RR Lyrae          &101& $-$214$\pm$10  & Dambis \& Rastorguev (\cite{D01}) \\
Subdwarf$^{b}$    & 97  & $-$208$\pm$6  & Carney \etal~(\cite{C96})       \\
Subdwarf$^{c}$    & 97  & $-$144$\pm$9  & Carney (\cite{C99})       \\
Subdwarf$^{d}$    & 230 & $-$161$\pm$7  & Chiba \& Beers (\cite{C00})  \\
\noalign{\smallskip}
\hline 
\noalign{\smallskip}
\end{tabular}           
\\
~~~$^{a}$~~~Stars with [Fe/H]$\leq$$-$1.6                     \\
~~~$^{b}$~~~Stars with [Fe/H]$\leq$$-$1.5 \& eccentricity $\leq$0.85  \\
~~~$^{c}$~~~Same as for Carney \etal~(\cite{C96}) but with kinematic bias
           correction  \\
~~~$^{d}$~~~Stars with [Fe/H]$\leq$$-$1.5                     \\
\end{flushleft}
\end{table*}
%----------------------------------------------------------- 

\begin{table*}
{{\bf Table~2.}The Halo rotation from stars out of the Galactic Plane } 
\begin{flushleft}
\begin{tabular}{lccc} 
\noalign{\smallskip}
\hline 
Halo tracer  & No. of stars& Mean Halo Rotation& Reference  \\
(Stellar type)& in sample   &  $<$V$>$ (km/s)       &            \\
\noalign{\smallskip}
\hline 
\noalign{\smallskip}
Subdwarf$^{a}$    & 21  & $-$275$\pm$16 & Majewski (\cite{M92}); \\
                  &     &               & Majewski \etal~(\cite{M96}) \\
Subdwarf$^{b}$    & 30  & $-$265$\pm$22 & Carney \etal~(\cite{C96})       \\
Subdwarf$^{c}$    & 30  & $-$196$\pm$13 & Carney (\cite{C99})       \\
Subdwarf$^{d}$    & 212 & $-$220$\pm$8  & Chiba \& Beers (\cite{C00})  \\
\noalign{\smallskip}
\hline 
\noalign{\smallskip}
\end{tabular}           
\\
~~~$^{a}$~~~{\it In situ} sample at North Galactic Pole       \\
~~~$^{b}$~~~Stars with [Fe/H]$\leq$$-$1.5 ; eccentricity $\leq$0.85 \& 
            Z$_{max}$ $\geq$ 4 kpc  \\
~~~$^{c}$~~~Same as for Carney \etal~(\cite{C96}) with Z$_{max}$ $\geq$ 4 kpc 
           and with kinematic ~~~~~~~~~ bias correction  \\
~~~$^{d}$~~~Stars with [Fe/H]$\leq$$-$1.5 \& Z$_{max}$ $\geq$ 4 kpc \\
\end{flushleft}
\end{table*}
%----------------------------------------------------------- 

\section{The Data}
\subsection{Selection of halo stars}
 We used blue horizontal branch (BHB) stars and RR Lyrae stars as tracers.
 The former  were selected from the surveys of Sanduleak (\cite{S88}) and
 Pesch \& Sanduleak (\cite{P89}) and 
 from the surveys of Beers \etal~(\cite {B96}). These candidate stars 
 were confirmed by $uBV$-photometry (Kinman \etal~\cite{K94}) and spectroscopy. 
 Most of the confirming spectra of the BHB stars were
 taken  at the Kitt Peak 4-m Mayall telescope (Kinman \etal~\cite {K96}). 
 RR Lyrae stars were selected
from the GCVS (Kholopov \cite{K85}) and subsequent Name-Lists and also from 
Kinman (\cite{K02a}). 
Intensity-weighted mean magnitudes of the RR Lyrae stars are derived from our 
recently observed light curves as these become available. 
 A program to obtain spectra of the RR Lyrae stars at the 3.5-m 
TNG telescope is in process. 
Only 6 of these RR Lyrae stars and none of the BHB 
stars in our current sample are included in the recent list of halo stars by
Beers \etal~(\cite{B00}).

\subsection{Absoute Magnitudes, Reddenings and Distances}

 The absolute magnitudes of RR Lyrae stars can be  determined 
from their metallicity [Fe/H] by a linear empirical relation of the form: 
 ~M$_{V}$ =  A [Fe/H] + B.  
 Chaboyer (\cite{C99}) gives 0.23 and 0.93  while Cacciari (\cite{C02}) 
 gives 0.23 and 0.92 respectively for A \& B. These values are
 consistent with an LMC modulus of 18.50 and give M$_{V}$ which are in the
 middle of the range of recent M$_{V}$ determinations (Popowski \& Gould 
\cite{P99}). [Fe/H], however,  has still to be determined for most of our
RR Lyrae sample, and so we used the relation based on the period (P) and 
Fourier coefficients (A1 and A3) given by Kov\'{a}cs \& Walker (\cite{K01}):
\begin{center}
     M$_{V}$ =  $-$1.876$\log$ P $-$ 1.158A1 + 0.821A3 + const. \\
\end{center}
The constant was taken as 0.448 (Kinman \cite{K02b}) which gives M$_{V}$ close
to the same scale as those derived from [Fe/H]. 
In the case of the BHB stars, we used the M$_{V}$ {\it vs.} $B-V$ relation 
given by Preston \etal~(\cite{P91}) adjusted  so that $M_{V}$ = +0.60 at 
$(B-V)$ = +0.20.  The reddening given by Schlegel \etal~(\cite{S98}) was adopted.
 
\subsection{Proper Motions}
We used proper motions that are based on the plate material used for the 
  construction of the GSC-II catalogue (Spagna \etal,~\cite{S96}). 
The relative proper motions are transformed to an absolute reference frame 
by forcing
the extended extragalactic sources to have null tangential motion. Since our
results depend critically on the success of this transformation, it seemed 
desirable to test the GSC-II proper motions of a sample of QSO which have an
image structure and colour similar to the program stars. 
The area studied contains 51 objects that have GSC-II proper motions and that 
are listed as QSO brighter than 18th magnitude 
by Hewitt \& Burbidge (\cite{H93}, HB) and V\'{e}ron-Cetty \&
V\'{e}ron (\cite{V01}, VV) in their catalogues. Six of these objects have
suprisingly large ($\geq$10 mas/yr) total proper motions and are listed in Table
3.  Spectra of these objects were taken by Arjun Dey and Buell Jannuzi using
the Kitt Peak 4.0-m telescope; their new classifications (private 
communication) are given in the final column of the Table. We note that VV
rejected CSO 832 and CSO 823 from their catalogue citing Sanduleak \& Pesch
(\cite{S90}) and Edwards \etal~(\cite{E88}) respectively. We therefore 
rejected the first five objects in Table 3 and assumed that the remaining 
46 objects are all QSO which should have no proper motion. These 46 objects 
have the following mean GSC-II proper motion: 
\begin{center}
 $\mu_{\alpha}$ = $-$ 0.646 $\pm$ 0.357  mas/yr   \\
 $\mu_{\delta}$ = $-$ 0.300 $\pm$ 0.507  mas/yr   \\
\end{center}
These mean proper motions are an indication of the systematic errors that 
could be present in the GSC-II proper motions over a sky area and magnitude
range similar to that of our program objects. In terms of the U and V 
velocity vectors at the North Galactic Pole, they correspond to:    
\begin{center}
 U = $-$1.8$\pm$1.9 km/s and V = $-$2.9$\pm$2.2 km/s at    1 kpc   \\
 U = $-$9.0$\pm$9.6 km/s and V = $-$14.3$\pm$11.0 km/s at    5 kpc  \\
 U = $-$18.0$\pm$19.2 km/s and V = $-$28.6$\pm$22.0 km/s at   10 kpc  \\
\end{center}  
This suggests that the systematic error in the mean rotation (V) is 
probably no more than 25 km/s (1$\sigma$ error) or at most 36 km/s 
(2$\sigma$ error) for our program objects with a mean distance of about
5 to 6 kpc.

\begin{table*}
{{\bf Table~3.} Objects described as QSO that have large GSC-II proper motions}
\begin{flushleft}
\begin{tabular}{lcccc} 
\noalign{\smallskip}
\hline 
Identification  & Catalogue & Redshift & Total proper &  Present     \\
                & Source    & Source   & motion (mas/yr) & Classification \\
\noalign{\smallskip}
\hline 
\noalign{\smallskip}
CSO~832              &  HB    &   V   & 24.3$\pm$2.7   &    Star     \\
CSO~823              &  HB    &   V   & 19.7$\pm$3.5   &    Star     \\
KUG~12491+2932       &  VV    &   D   & 18.5$\pm$2.5   &    Star     \\
CSO~835              &  HB,VV &   V   & 16.8$\pm$3.9   &    Star     \\
1306+293; BG 57~34   &  HB,VV &   V   & 13.5$\pm$3.1   &    Star     \\
1306+274             &  HB,VV &   W   & 10.2$\pm$5.4   &    QSO      \\
\noalign{\smallskip}
\hline 
\noalign{\smallskip}
\end{tabular}           
\\
 Redshift references:~~~ V = Vaucher (\cite{V82}); D = Darling
 \etal~(\cite{D94}); \\
~~~~~~~~~~~~~~~~~~~~~~~~~~~~~ W = Wills \& Wills (\cite{W76})     \\
\end{flushleft}
\end{table*}
%----------------------------------------------------------- 

\subsection{Results} Our complete sample consists of 87 confirmed BHB stars 
that have b $\geq$ 75$^{\circ}$ (72 of which have b $\geq$ 80$^{\circ}$). There 
are also 73 RR Lyrae stars in this area for which we are in the process of 
getting light curves and spectra. Our present results refer to a subset of 35 
BHB stars (30 having radial velocities) and 18 RR Lyrae stars (of which 9 have 
radial velocities). In calculating the UVW vectors (Johnson \cite{J87}), 
we put the radial velocity equal to zero (with an error of $\pm$150 km/s) if no 
radial velocity was available. In such cases, the U and V vectors should be 
very close to the correct values 
but the W velocity  must be discarded. These heliocentric UVW are 
compared with those found by Martin \& Morrison (\cite{M98}) for their HALO2
sample of local RR Lyrae stars. They excluded disk RR Lyrae stars and 
trimmed 10\% of the ``outlyers" from their sample. We have not attempted to 
remove disk stars from our sample (Z $>$ 1.6 kpc) but found that
trimming hardly changes the mean values of U, V and W although it does
reduce the velocity dispersions $\sigma_{u}$, $\sigma_{v}$ and $\sigma_{w}$. 
Martin \& Morrison used a M$_{V}$
that is less than 0.1 mag fainter than ours, but this should not account for
the 60 km/s difference in V velocities (but very comparable $\sigma_{v}$). 
The 42 stars in our sample with Z $<$ 10 kpc have almost the same $<$V$>$ 
($-$286$\pm$19 km/s at $<$Z$>$ = 5.3 kpc) as for the whole sample. 

\begin{table*}
{{\bf Table~4.} UVW velocities for our NGP sample compared with those of
Martin \& Morrison (\cite{M98})}
\begin{flushleft}
\begin{tabular}{lccccccc} 
\noalign{\smallskip}
\hline 
  U    &   V   &   W  &$\sigma_{u}$&$\sigma_{v}$&$\sigma_{w}$& No. of & Source \\
km/s  &km/s & km/s  & km/s & km/s & km/s & stars  &                         \\
\noalign{\smallskip}
\hline 
\noalign{\smallskip}
$-$7$\pm$22 &$-$285$\pm$17 & $-$26$\pm$15 & 155 & 125 & 92 & 53 (39) & (1)  \\
$-$1$\pm$19 &$-$285$\pm$14 & $-$25$\pm$13 & 127 &  92 & 76 & 47 (35) & (2)  \\
$-$1$\pm$26 &$-$219$\pm$24 & $-$5$\pm$10 & 193 & 91 & 96 & 84  & (3)  \\
\noalign{\smallskip}
\hline 
\noalign{\smallskip}
\end{tabular}           
\\
Sources :~~~ (1) Present paper; (2) Present sample (10 \% outlyers trimmed); \\
~~~~~~~~~~~~~~~ (3) Martin \& Morrison (\cite{M98})     \\
\end{flushleft}
\end{table*}
Gilmore \etal~(\cite{G02}) have recently reported retrograde halo rotation 
out of the Plane from their radial velocity determinations of stars at 
galactic longitude 270$^{\circ}$. We shall determine UVW from our total 
sample at the North Galactic Pole in the near future, and hope to extend 
the work to halo stars in Anticentre fields. 
Possibly this will allow us to detect 
gradients in the V motion and discover whether this retrograde rotation is 
caused by local streaming or is part of a more widespread effect. 
%----------------------------------------------------------- 

%%-----------------------------
%%      your bibliography
%%-----------------------------

\end{document}